# First-principles calculation on the electronic structures, phonon dynamics, and electrical conductivities of $Pb_{10}(PO_4)_6O$ and $Pb_9Cu(PO_4)_6O$ compounds

*L. Y. Hao, E. G. Fu\**

State Key Laboratory of Nuclear Physics and Technology, Department of Technical Physics, School of Physics, Peking University, Beijing 100871, P. R. China

**Abstract:** Superconducting materials with high critical temperature have the potential to revolutionize many fields, including military, electronic communications, and power energy. Therefore, Scientists around the world have been tirelessly working with the ultimate goal of achieving high temperature superconductivity. In 2023, a preprint by S. Lee et al in South Korea claimed the discovery of ultra-high-temperature superconductivity with a critical temperature of up to 423 K in Cu-doped lead-apatite (LK-99) (arXiv:2307.12008, arXiv:2307.12037), which caused a worldwide sensation and attention. Herein, the electronic structures, phonon dynamics, and electrical conductivities of LK-99 and its parents compound lead-apatite have been calculated using first-principles methods. The results show that the lead-apatite compound and the LK-99 compound are insulator and half-metal respectively. The flat band characteristic is consistent with previous calculations. The electrical conductivity of LK-99 compound shows two extreme point, and the electrical conductivity along the C-axis increases significantly after 400 K. The phonon dispersion spectra of the compounds were investigated, demonstrating their dynamic instability.

**Keywords:** Superconductivity, LK-99, Lead-apatite, First-principles calculation

\* Corresponding author

E-mail addresses: efu@pku.edu.cn (E.G. Fu).

## 1. Introduction

A superconductor is a material that exhibits zero electrical resistance below a certain critical temperature [1] and has three fundamental properties: superconductivity [2], complete diamagnetism (Meissner effect) [3,4], and magnetic flux quantization [5,6]. Superconductors can be classified into two types based on their response to magnetic fields: Type I superconductors [7] have a single critical magnetic field, above which superconductivity is lost, while Type II superconductors [8] have two critical magnetic field values, between which the material allows partial penetration of the magnetic field. Superconducting materials with high critical temperature have the potential to revolutionize many fields, including military [9], electronic communications [10], and power energy [11]. In particular, superconductors are essential for controlled nuclear fusion [12], as the ultra-high temperature plasma in the fusion core must be confined by a large magnetic field, which can only be provided by superconducting materials [13]. However, current superconductors must operate at extremely low temperatures, which requires significant energy for cooling [14]. The development of room-temperature superconductors would greatly increase the efficiency of nuclear fusion reactors and could potentially solve global energy problems [15]. Since the discovery of superconductivity in 1911 [16], scientists around the world have been tirelessly working with the ultimate goal of achieving high temperature superconductivity, but most superconductors only work at extremely low temperatures (less than 77K) [17].

Currently, the main types of high-temperature superconductors include cuprate, iron-based [18], organic [19], and hydride superconductors [20,21]. In 2015, a hydrogen sulfide superconductor was discovered by the research groups of V. Ksenofontov and S. I. Shylin et al. at the Max Planck Institute, with a record-breaking critical temperature of 203 K [22]. In 2019, M. I. Eremets et al. [23] from Germany successfully prepared $LaH_{10}$ and demonstrated its superconducting properties at 250 K. This research marked a transition from the discovery of superconductors by trial and error to the use of theoretical predictions to guide research [24]. In 2023, an arXiv preprint by S. Lee et al [25,26] in South Korea claimed the discovery of ultra-high-temperature superconductivity with a critical temperature of up to 423 K in Cu-doped lead-apatite (LK-99), with the chemical formula of $Pb_{10-x}Cu_x(PO_4)_6O$ (0.9<x<1.1), and provided detailed methods for sample preparation. However, the superconductivity has not yet been verified and the sample size is too small. First-principles calculations are often used to guide the search for new superconductors, and now many studies have investigated the electronic structures of LK-99 [27-29], but its electrical conductivity and phonon dispersion properties have not been properly addressed.

Herein, the electronic structures, phonon dynamics, and electrical conductivities of the $Pb_{10}(PO_4)_6O$ and $Pb_9Cu(PO_4)_6O$ compounds have been calculated using first-principles methods. The results show that the $Pb_{10}(PO_4)_6O$ compound is an insulator, while the $Pb_9Cu(PO_4)_6O$ compound exhibits half-metallic behavior. The flat band characteristic is consistent with previous calculations [27]. The temperature-dependent conductivity curves of the two compounds were also calculated and analyzed, revealing novel properties of their conductivity. Additionally, the phonon dispersion spectra of the compounds were investigated, demonstrating their dynamic instability. $Pb_9Cu(PO_4)_6O$ compound may have novel magnetic properties and electrical transport properties, but the preparation of large size $Pb_9Cu(PO_4)_6O$ compound for experimental research is still a difficult issue.

2. **Computational details**

The first-principles calculations in this work are done by using the projected plane wave (PAW) method [30] within the Vienna Ab initio Simulation Package (VASP) [31-33]. The hexagonal structure of $Pb_{10}(PO_4)_6O$ compound, which is belong to P63/m space group [34], was constructed with the help of VESTA program [35]. Considering the symmetry of the system, 3 oxygen atoms at the vertices sites are 1/4 occupied and were removed as described in the literature [27]. The system was geometrically optimized firstly and then the electronic and dynamical properties were systematically calculated on the energy-stable structure. The Perdew-Burke-Ernzerhof (PBE) functional in Generalized Gradient Approximation (GGA) [36] was set to describe the exchange correlation energy. Based on the convergence test, a plane wave cut-off energy of 520 eV and a 3×3×3 k-mesh of Monkhorst-Pack grid for the first Brillouin zone was selected. The self-consistent convergence tolerance was set to 1×10$^{-6}$ eV/atom, and the maximum ionic force tolerance was set to 0.01 eV/Å. Considering the possible magnetic and spin polarization properties of Cu atom, the spin polarization in all calculations was switched on when calculating the properties of Pb$_9$Cu(PO$_4$)$_6$O compound, which is contrastively switched off in that of Pb$_{10}$(PO$_4$)$_6$O compound. The calculation of the electronic band structures and density of states (DOS) were done with post-processing VASPKIT package[37]. The phonon dispersion spectrums were calculated using the PHONOPY package [38,39].

## 3. Results and discussion

There are two different kinds of Pb in Pb$_{10}$(PO$_4$)$_6$O compound, which occupy the lattice points a large hexagon and two triangles with a 180° angle to each other. According to previous literature [27], Cu atoms tend to displace Pb atoms at the lattice points of large hexagons when forming Pb$_9$Cu(PO$_4$)$_6$O compound. Based on these, the structural configurations of both two compounds were constructed as shown in Figure 1(a) and Figure 2(a). The lattice constants and cell volumes of Pb$_{10}$(PO$_4$)$_6$O compound and Pb$_9$Cu(PO$_4$)$_6$O compound after structural relaxation are shown in Table 1. For these two compounds, the lattice constant a is equal to b, and the lattice angle α=β=90°, γ=120°. Our work is in good agreement with previous experimental results as shown in Table 1. The relative error in lattice constants of Pb$_{10}$(PO$_4$)$_6$O compound is only about 0.2%, while the relative error of Pb$_9$Cu(PO$_4$)$_6$O compound is about 0.4%. This degree of error is definitely acceptable for first-principles calculations.

Table 1. The calculated lattice constants, cell volumes and band gaps obtained by first-principles calculation, which are compared to the experimental results.

| Structures | a=b (Å) | c (Å) | Volume (Å$^3$) | Band gap |
| --- | --- | --- | --- | --- |
| Pb$_{10}$(PO$_4$)$_6$O | 9.888 | 7.438 | 642.75 | 2.67 eV |
| Exp. [40] | 9.865 | 7.431 | 626.25 | Insulator |
| Pb$_9$Cu(PO$_4$)$_6$O | 9.802 | 7.339 | 610.75 | Half metal |
| Exp. [26] | 9.843 | 7.428 | 623.24 | Metal |

Then the electronic structural properties of both compounds were investigated under the equilibrium structural configurations, including the charge densities, band structures and DOS. Figure 1(b) shows the charge density of Pb$_{10}$(PO$_4$)$_6$O compound. It can be clear seen that the metal atoms and oxygen atoms are connected by covalent bonds, and the charge is mostly concentrated on the oxygen side rather than the Pb and P sides. The calculated band structure and DOS of Pb$_{10}$(PO$_4$)$_6$O compound are plotted in Figure 1(c). The compound is non-magnetic, because all the atoms inside have near zero net magnetic moment. Therefore, it also has no spin polarization properties. According to our calculations, the band gap of Pb$_{10}$(PO$_4$)$_6$O compound is 2.67 eV, which was also shown in Table 1, and such a large bandgap are usually the characteristic of insulators or wide bandgap semiconductors. For the sake of demonstration, we set the Fermi level at E=0 eV. It is noted that below the Fermi level, the electron state density is mainly provided by the 2p electrons of the O atoms, while above the Fermi level, the electron state density is mainly provided by the 6s6p electrons of the Pb atoms. Two bands in the valence band close to the Fermi level are marked in red because they are obvious flat band which is consistent with the calculation results in the previous work [27]. It can be seen that the band structure is flat and near the Fermi surface in the L→H region, which means that the energy dispersion of the electrons here is almost independent of the change in momentum. In addition, the literally small slope of the band structure curves means the localization of the electronic system is very strong, and this kind of strongly correlated system may have a novel physical phenomenon under certain conditions, and superconductivity is one of them. The DOS spectrums show that the two energy bands near the Fermi surface are mainly provided by the O-2p electrons and the Pb 6s6p electrons, and the contributions of both are indispensable.

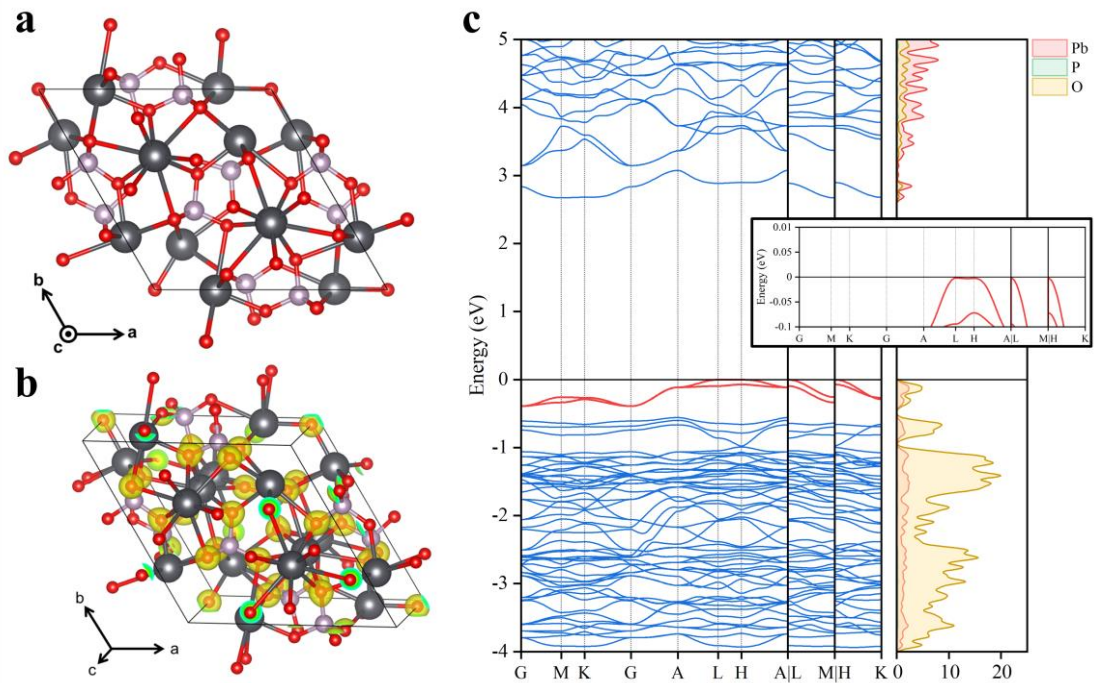

Figure 1. (a) The structural configuration of $Pb_{10}(PO_4)_6O$ compound. (b) The charge density distribution of $Pb_{10}(PO_4)_6O$ compound. (c) The band structure and DOS spectrum of $Pb_{10}(PO_4)_6O$ compound.

Things are different when it comes to $Pb_9Cu(PO_4)_6O$ compound. It's noted that the lattice parameters are reduced by about 0.87% and the lattice volume is shrunk by 4.97% with the introduction of Cu atom, which accords with the law of experimental observation [26,40]. The system shows net magnetic moment of 0.56 $\mu_B$, all of which is provided by Cu atom, and spin-polarized property and may possess novel characteristics in electricity and magnetism. Figure 2(b) shows the charge density distribution of $Pb_9Cu(PO_4)_6O$ compound, indicates that the charge is obviously concentrated near the Cu atom except the O atoms, which is different from the $Pb_{10}(PO_4)_6O$ compound and may contributed to special magnetic properties at macroscopic scale. The calculated band structure and DOS of $Pb_9Cu(PO_4)_6O$ compound are shown in Figure 2(c). Taking spin polarization into account, we consider the electrons with different spin directions separately. In the band diagram, the electrons with spin-up are represented by the solid blue line, while the electrons with spin-down are represented by the dashed red line. In the DOS spectrum, the spin-up electron contribution has a positive state density, while the spin-down electron contribution has a negative state density. The substitution Cu atom introduced two defect levels in the band gap of $Pb_{10}(PO_4)_6O$ compound, which, miraculously, happened to be located at the Fermi energy level, giving the entire structure half-metallic properties to some extent. The two defect levels are flat bands with energy dispersion ranges of less than 0.1 eV in the full momentum space, and both the upper one and the lower one is a half-filled band, which shows differences from conclusions in the previous work [29]. In the same way, we expect this strongly correlated system to bring about novel physical properties. But it is worth noting that the flat band is only for spin-up electrons, while for spin-down electrons, the band still behaves as an insulator with a band gap of about 3 eV. The DOS spectrum shows that the flat band is mainly contributed by O-2p electrons and Cu-3d electrons, which are both have spin-up direction. This proves that only structures in which Cu atom replace Pb atom conduct electricity, and the others are insulators. It means that the oxygen atoms enclosed inside the two triangles (in other words, the oxygen atom on the lattice point of unit cell) are confined to their one-dimensional channels along the C-axis [27], as shown in the diagram in Figure 3.

Based on this, the conductivity of the material along different coaxial directions was calculated according to the semi-empirical Boltzmann transport theory. The group velocity of an electron can be expressed by the following formula [41]:

$$v_\alpha(i,k) = \frac{1}{h}\frac{\partial \varepsilon_{i,k}}{\partial k_\alpha}$$

$$v_\beta(i,k) = \frac{1}{h}\frac{\partial \varepsilon_{i,k}}{\partial k_\beta}$$

where $\varepsilon_{i,k}$ is the energy eigenvalue of band No. $i$ at point $k$; $\alpha$ and $\beta$ means different electron velocity directions. The conductivity tensor can therefore be expressed as [41]:

$$\sigma_{\alpha\beta}(i,k) = e^2 \tau_{i,k} v_\alpha(i,k) v_\beta(i,k)$$

where $\tau_{i,k}$ is the relaxation time, which can be treated as a constant in most cases without affecting the accuracy of the calculation. However, $Pb_9Cu(PO_4)_6O$ compound is different from common metals or insulators, so its relaxation time constant has not been estimated here, and its electrical conductivity was only discussed qualitatively.

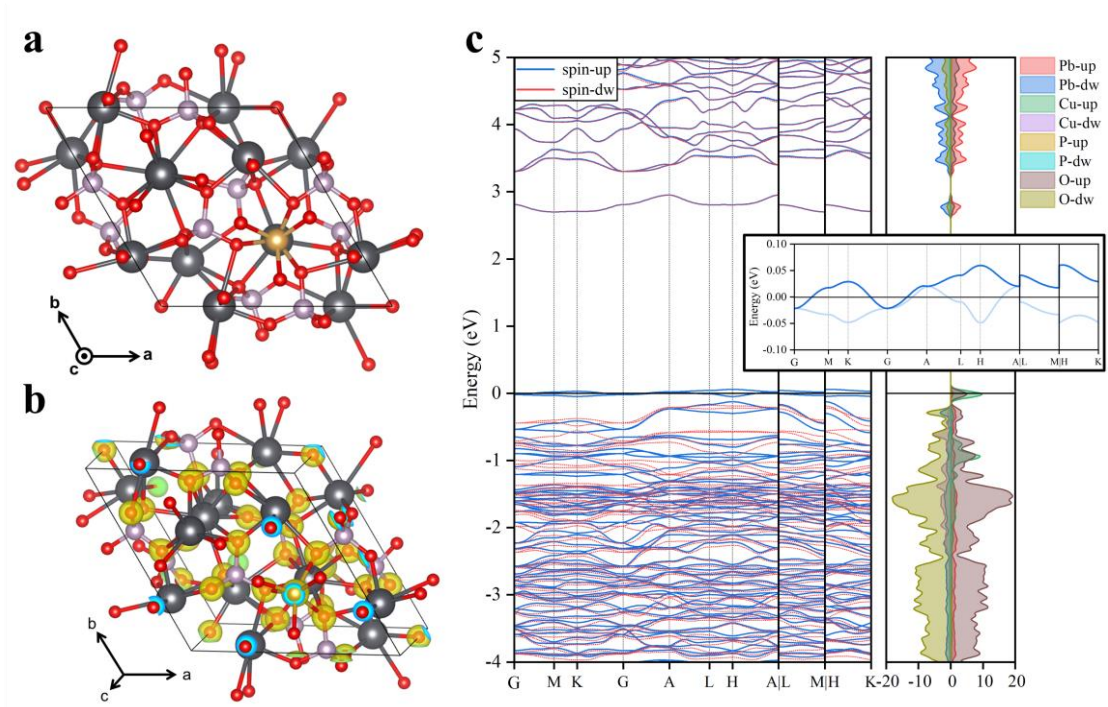

Figure 2. (a) The structural configuration of $Pb_{10}(PO_4)_6O$ compound. (b) The charge density distribution of $Pb_9Cu(PO_4)_6O$ compound. (c) The band structure and DOS spectrum of $Pb_9Cu(PO_4)_6O$ compound.

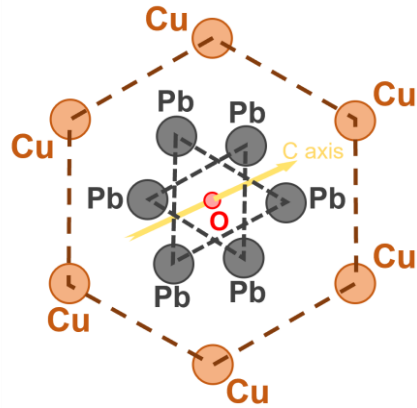

Figure 3. Schematic diagram of a one-dimensional conductive channel for O atoms.

We defined the generalized electrical conductivity as the electrical conductivity in units' relaxation time. Figure 4(a) and 4(b) respectively show the generalized electrical conductivity curves of $Pb_{10}(PO_4)_6O$ compound and $Pb_9Cu(PO_4)_6O$ compound with temperature variations ranging from 100 K to 450 K. All three axes are considered, where XX represents A-axis, YY represents B-axis, ZZ represents C-axis, and the average conductivity is provided for reference. It is evident from the figures that there is no significant difference in the electrical conductivity of these two compounds in A-axis and B-axis, so their curves coincide. For $Pb_{10}(PO_4)_6O$ compound, the electrical conductivity of the three axes is close to 0 at low temperature. With the increase of temperature, the carrier inside the material is excited, and its electrically conductive ability is improved, which is in line with the characteristics of its insulator. The increase of the electrical conductivity of $Pb_{10}(PO_4)_6O$ compound in the direction of C-axis with temperature is obviously greater than that in the direction of A-axis and B-axis, and it even reaches about 20 times of the latter at 450 K.

Besides, the conductivity curve of $Pb_9Cu(PO_4)_6O$ compound is very interesting. Firstly, its conductivity decreases with increasing temperature, which is clearly a metallic property, and its value is two orders of magnitude higher than that of $Pb_{10}(PO_4)_6O$ compound. Secondly, the conductivity along its A-axis and B-axis has a maximum value near 120 K, while the conductivity along its C-axis has a minimum value near 380 K. This results in a maximum and a minimum of the average conductivity of the material at about 110 K and 420 K respectively. According to the calculation results of the energy band and DOS, $Pb_9Cu(PO_4)_6O$ compound should be a metal, but the electrical conductivity of the metal generally decreases with the increase of temperature. Amazingly, its conductivity conforms to the general law of metals only on the A-axis and B-axis, while the conductivity along the C-axis increases significantly after 400 K, which we speculate is related to the one-dimensional channel along the C-axis of the 1/4 occupied O atoms surrounded by Cu atoms and Pb atoms (see Figure 3), but there is no further verification at present.

Unfortunately, our calculation results do not support the superconducting properties of $Pb_9Cu(PO_4)_6O$ compound in the temperature range of 100-450 K. Some typical metals such as Fe, Cu, Hf, were calculated, and finally found their electrical conductivity in unit relaxation time is 3-4 orders of magnitude larger than that of $Pb_9Cu(PO_4)_6O$ compound, even along the C-axis, but it is not ruled out that the electronic system of this novel structure cannot be simply explained by the existing theory, so we reserve our views on its superconducting properties.

Considering the spin polarization characteristics of $Pb_9Cu(PO_4)_6O$ compound, the generalized electrical conductivity of different spin-channels has been calculated respectively, and the results are shown in Figure 4(c) and 4(d), where the former is the conductivity for the spin-up electrons in $Pb_9Cu(PO_4)_6O$ compound, and the latter one is for the spin-down electrons. It can be easily found that its spin up electrons has much greater contribution to electrical conductivity than the spin-down electrons. This is consistent with the band structures calculation results: the spin-up channel is metallic, while the spin-down channel is insulating. Therefore, the total electrical conductivity curve is basically the same as the electrical conductivity curve of the spin-up electrons.

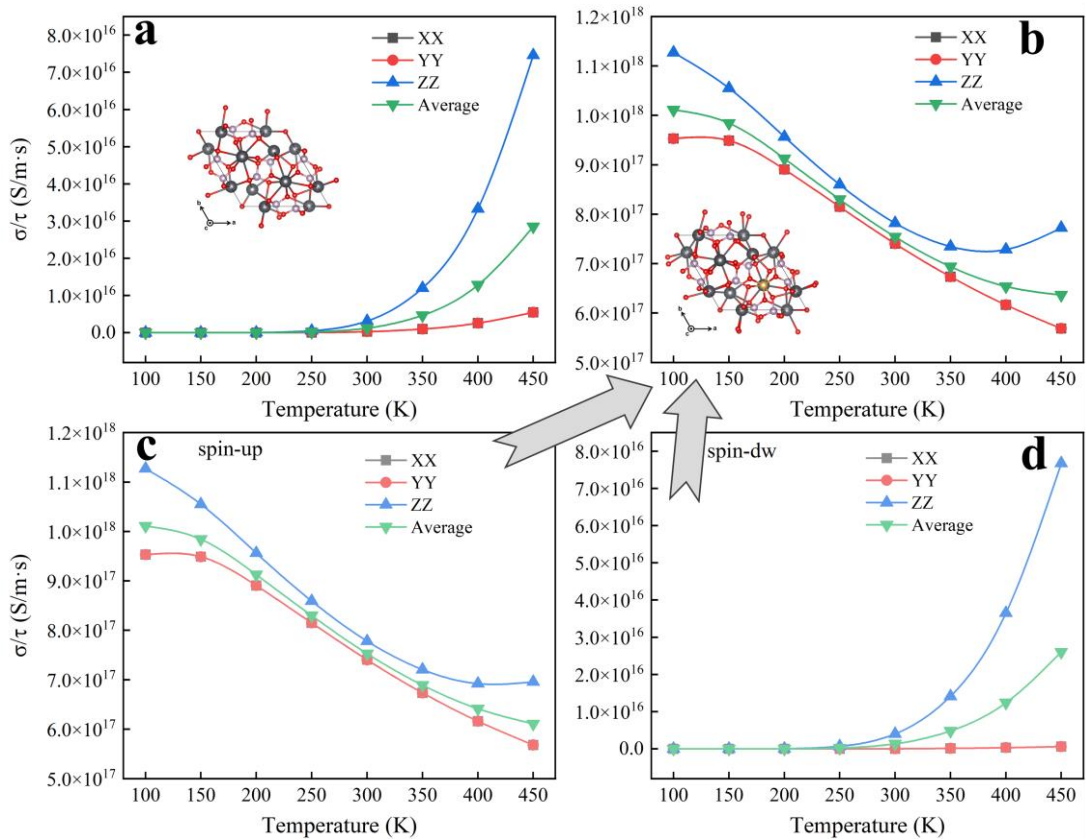

Figure 4. (a) Generalized conductivity curve of $Pb_{10}(PO_4)_6O$ compound with temperature. (b) Generalized conductivity curve of $Pb_9Cu(PO_4)_6O$ compound with temperature. (c) Generalized spin-up conductivity curve of $Pb_9Cu(PO_4)_6O$ compound. (d) Generalized spin-down conductivity curve of $Pb_9Cu(PO_4)_6O$ compound.

Considering the difficulty in laboratory synthesis of $Pb_{10}(PO_4)_6O$ compound and $Pb_9Cu(PO_4)_6O$ compound, we hypothesized that these two structures may not be dynamically stable. The calculated phonon spectra of these two compounds are shown in Figure 5. Several phonon spectra of $Pb_{10}(PO_4)_6O$ compound and $Pb_9Cu(PO_4)_6O$ compound extend to negative values, while $Pb_9Cu(PO_4)_6O$ compound even has two vibratory spectral lines starting below -2 THz. This proves that these structures, especially $Pb_9Cu(PO_4)_6O$ compound, may not be the optimal solution at the energy scale. It may need to be synthesized at very small scales because the additional interface or surface energy can support for the formation of these two phases. Above these results, our conjecture that they are dynamically unstable structures has been confirmed, and the synthesis process of $Pb_9Cu(PO_4)_6O$ compound is more difficult.

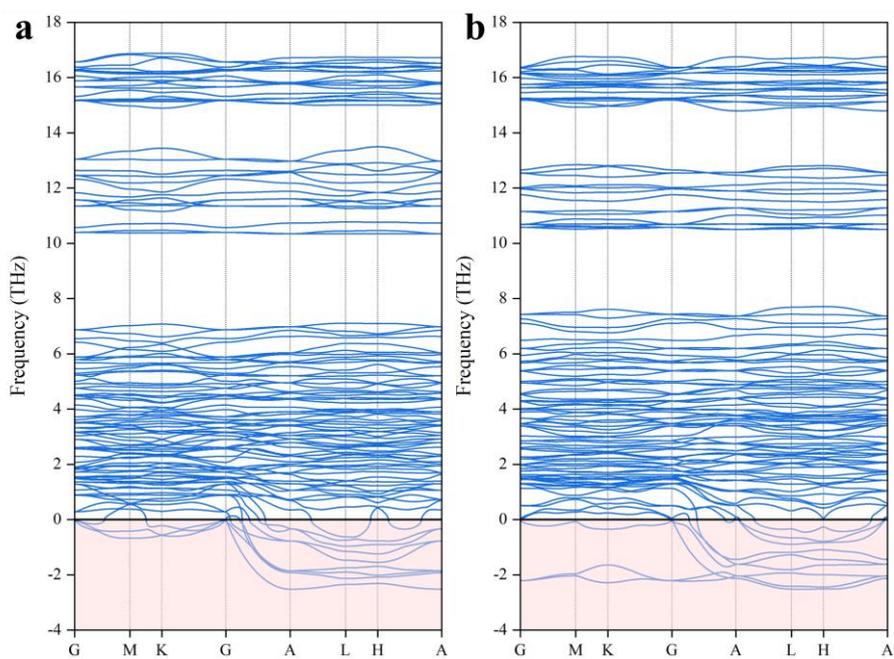

Figure 5. (a) Phonon dispersion spectra of $Pb_{10}(PO_4)_6O$ compound. (b) Phonon dispersion spectra of $Pb_9Cu(PO_4)_6O$ compound.

## 4. Conclusion

In summary, the electronic structural properties, phonon dynamic properties, and electrical conductivities of $Pb_{10}(PO_4)_6O$ compound and $Pb_9Cu(PO_4)_6O$ compound have been calculated and analyzed by first-principles calculations. Band and DOS calculation results show that $Pb_{10}(PO_4)_6O$ compound is an insulator, while $Pb_9Cu(PO_4)_6O$ compound exhibits half-metallic properties, which is insulator for spin-down electrons but metallic for spin-up electrons, and has two half-filled flat bands at the Fermi energy level, which are produced by the interaction of O-2p electrons and Cu-3d electrons. The calculated generalized electrical conductivity results are in good agreement with those of band structures and DOSs, but the electrical conductivity of $Pb_9Cu(PO_4)_6O$ compound shows two extreme point, and the electrical conductivity along the C-axis increases significantly after 400 K, which is possibly related to the one-dimensional channel along the C-axis of the 1/4 occupied O atoms surrounded by Cu atoms and Pb atoms. Unfortunately, our results show that the electrical conductivities in unit relaxation time of some typical metals are 3-4 orders of magnitude greater than that of $Pb_9Cu(PO_4)_6O$ compound, which seems difficult to support the high temperature superconductivity in it. The results of phonon dispersion spectra show that both compounds are dynamically unstable, and $Pb_9Cu(PO_4)_6O$ compound is more difficult to synthesize in the laboratory. Its superconductivity has not observed in the experiment so far, and the mechanism of observed diamagnetism also connot be explained by the calculated magnetic moments, which still needs to be further explored. However, no matter whether it is subsequently confirmed as a room-temperature superconductor, the preparation of large size $Pb_9Cu(PO_4)_6O$ compound for experimental research is still a difficult issue.

## Acknowledgement

This work was supported by the National Natural Science Foundation of China (11921006, 11975034, 11375018, 12005048, U20B2025 and U21B2082), Beijing Municipal Natural Science Foundation (1222023), Shanghai Sailing Program (20YF1453900), and Natural Science Foundation of Shenzhen (GXWD20201230155427003-20200822141815001). This work is also supported by High-performance Computing Platform of Peking University. E.G. Fu acknowledges the support from Science Fund for Creative Research Groups of NSFC, the Ion Beam Materials Laboratory (IBML) at Peking University.

## Declaration of Competing Interest

The authors declare that they have no known competing financial interests or personal relationships to disclose.